\documentclass[%
reprint,
superscriptaddress,
nofootinbib,
amsmath,amssymb,
]{revtex4-1}

\pdfoutput=1
\usepackage{float}
\usepackage{xcolor}
\usepackage{graphicx}
\usepackage{dcolumn}
\usepackage{bm,lipsum}
\usepackage{siunitx}
\usepackage[colorlinks=true, linkcolor=blue, citecolor=blue, urlcolor=blue]{hyperref}
\newcommand{\ZIB}{Zuse Institute Berlin, 14195 Berlin, Germany}
\newcommand{\JCM}{JCMwave GmbH, 14050 Berlin, Germany}

\usepackage{graphicx}
\usepackage{booktabs}
\usepackage{tabularx}
\usepackage{xcolor,colortbl}
\usepackage[T1]{fontenc}

\begin{document}

\title{Efficient rational approximation of optical response functions \\ with the AAA algorithm}
\author{Fridtjof Betz}
\affiliation{\ZIB}
\author{Martin Hammerschmidt}
\affiliation{\JCM}
\author{Lin Zschiedrich}
\affiliation{\JCM}
\author{Sven Burger}
\affiliation{\ZIB}
\affiliation{\JCM}
\author{Felix Binkowski}
\affiliation{\ZIB}

\begin{abstract}
We introduce a theoretical framework for the rational approximation of optical response functions in resonant photonic systems. The framework is based on the AAA algorithm and further allows to solve the underlying nonlinear eigenproblems and to efficiently model sensitivities. An adaptive sampling strategy exploits the predominance of resonances in the physical response. We investigate a chiral metasurface and show that the chiroptical response on parameter variations can be accurately modeled in the vicinity of the relevant resonance frequencies.  
\end{abstract}

\maketitle
Resonances are substantial for the functionality of photonic devices.
For example, single molecule detectors with ultrahigh sensitivity~\cite{Nie_Science_1997,Anker_BioSens_NatMater_2008}, 
efficient single-photon sources~\cite{Reithmaier_2004,Senellart_2017}, and devices for photocatalysis~\cite{Ma_LiSciAppl_2016,Zhang_ChemRev_2018} are based on resonance effects. The devices are typically analyzed by measuring the electromagnetic response in a real-valued frequency range, where the shapes of the corresponding optical response functions are characterized by the underlying resonances~\cite{Novotny_Hecht_2012}. Numerical simulations for the real-valued excitation frequencies are performed to understand and optimize the devices.

However, the response of the corresponding non-Hermitian systems can be locally approximated by a rational function of the excitation frequency with poles at the complex-valued resonance frequencies~\cite{Lalanne_QNMReview_2018,Zworski_Scattering_Resonances_2019}. Therefore, exploiting the complex frequency plane is a natural approach for the investigation of the systems. The source-free Maxwell's equations can be solved to obtain the resonance modes and resonance frequencies~\cite{Lalanne_QNM_Benchmark_2018,Demesy_ComputPhysComm_2020}. Subsequently, the modes are used for a reconstruction of the optical response functions on the real frequency axis~\cite{Sauvan_2022,Nicolet_2022}. The functions are decomposed into weighted sums of modes. This makes it possible to understand which resonances are relevant for the systems. However, no physical information regarding the response functions of interest and their excitation sources is used when the resonances are computed. Therefore, it can be necessary to compute many resonances which have negligible contribution to the reconstruction~\cite{Yan_PRB_2018}. A direct rational approximation of the response functions using information from a discrete set of sample points in the complex frequency plane can also provide the required information on the real frequency axis. The response functions are physical observables, i.e., physical information is directly incorporated into the approximation. This means that only the resonance frequencies that are relevant for the physical quantity under consideration are taken into account in the rational approximation~\cite{Binkowski_PRB_2024}.

In photonics, not only the resonances and the response on the real frequency axis, but also their sensitivities with respect to the system parameters are of fundamental interest~\cite{Jensen_LasPhotRev_2011}. The sensitivities of the resonances can give a better understanding of the systems and they can be used for gradient-based optimization schemes~\cite{Swillam_2008,Burschaepers_2011,Yan_2020,Binkowski_CommunPhys_2022}.
Therefore, there is a need for theoretical approaches providing all the required quantities together, i.e., resonance modes and resonance frequencies, reconstruction on the real frequency axis, and sensitivities. Numerical implementations of such approaches should be efficient, accurate, straightforward to implement, and stable when applied to complex physical setups. 

\begin{figure}
\includegraphics[width=0.49\textwidth]{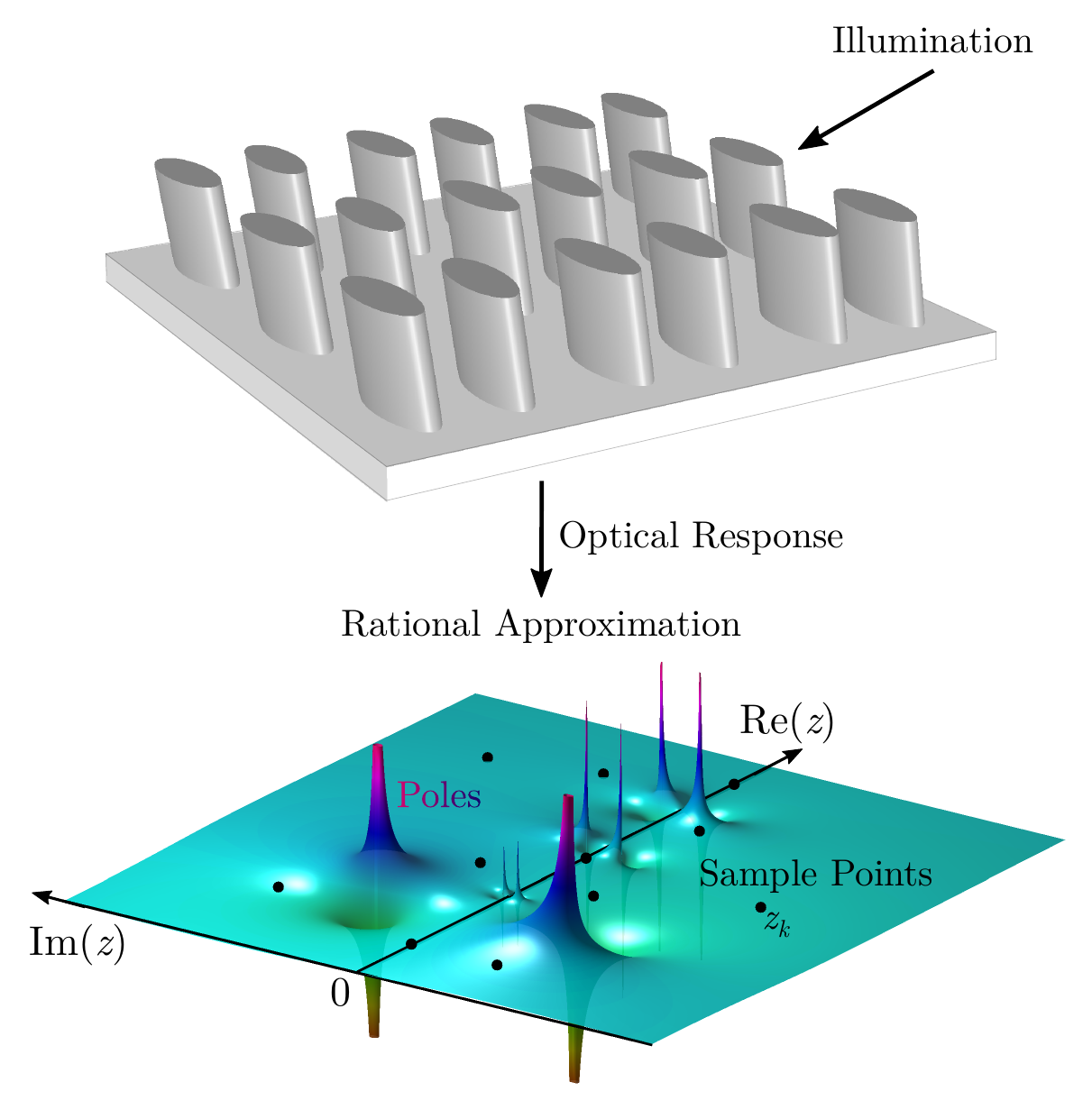}
\caption{\label{fig1}
Rational approximation of optical response functions. Illumination of a metasurface leads to an optical response. The optical response is characterized by the underlying complex-valued resonance frequencies and can be approximated by a rational function. The poles of the rational approximation are located at the resonance frequencies. The approximation is based on the optical response at adaptively selected sample points in the complex frequency plane.
}
\end{figure}

In this work, we present a theoretical approach for the investigation of resonant photonic systems. The approach is based on rational approximation using the AAA algorithm. The AAA algorithm has recently been introduced~\cite{Nakatsukasa_2018_AAA} and applied to a large number of problem classes in physical sciences~\cite{Filip_2018,Guettel_2022,Trefethen_2023,Keith_JFluidMech_AAA_2021,Kehry_JChemPhys_AAA_2023}. We use the AAA algorithm to obtain approximations of optical response functions in the complex frequency plane, including the real axis. We further introduce an approach for solving the underlying nonlinear eigenproblems and for rational approximation of sensitivities. Accurate results are obtained by applying the rational approximation directly to the physical response of interest and by an adaptive sampling strategy. We use the presented framework to investigate optical properties of a chiral metasurface. 
Figure~\ref{fig1} shows a schematics of the investigation. The transmission through the illuminated metasurface is the optical response of interest. It is approximated by a rational function with poles at the complex-valued resonance frequencies. The rational function is built based on the response at adaptively chosen sample points in the complex frequency plane.

\emph{Rational approximation with the Adaptive Antoulas--Anderson (AAA) algorithm}.---Recently, Nakatsukasa, Sète, and Trefethen introduced the AAA algorithm~\cite{Nakatsukasa_2018_AAA}. A scalar function $f(z)$ is approximated by a rational function $r(z)$  
on a finite domain in the complex plane. Typically, $M$ unique sample points $z_k$ from a finite set $Z\subseteq\mathbb{C}$ along with the corresponding function values $f_k = f(z_k)$ are provided. The algorithm greedily adds points $\hat{z}_j$ to a subset $\hat{Z}\subset Z$. Together with the corresponding function values $\hat{f}_j$, each iteration leads to a rational approximation $r(z)$ of order $m-1$ in barycentric representation,
\begin{equation} \label{eq:r}
    r(z) = \frac{n(z)}{d(z)} = \left. \sum_{j=1}^{m}\frac{\hat{w}_j \hat f_j}{z-\hat z_j}\middle/\sum_{j=1}^{m} \frac{\hat{w}_j}{z-\hat z_j} \right.,
\end{equation}
with weights $\hat{w}_j$ that minimize the error 
\begin{equation} \label{eq:lsq}
    \sum_{z_k\in Z\backslash\hat{Z}}|f_k \, d(z_k)-n(z_k)|^2
\end{equation} 
under the constraint
\begin{equation} \label{eq:nrm}
    \sum_{j=1}^m |\hat{w}_j|^2 = 1.
\end{equation}
The zeros of $n(z)$ and $d(z)$ are the zeros and poles of $r(z)$, respectively. They are provided as the results of generalized eigenproblems, and the least squares problem in Eq.~\eqref{eq:lsq} is solved with a singular value decomposition~\cite{Nakatsukasa_2018_AAA}.

The barycentric form introduced in Eq.~\eqref{eq:r} has removable singularities at $\hat{z}_j \in \hat{Z}$ and the limit $\lim_{z\rightarrow\hat{z}_j}r(z) = \hat{f}_j$ exists. Therefore, the approximation $r(z)$ interpolates the function values $\hat{f}_j$ and, to define a proper least square problem, Eq.~\eqref{eq:lsq} requires $m\leq M/2$. We will use the remaining support points to accurately approximate the sensitivities of poles $z_n^\mathrm{poles}$, corresponding residues $a_n$, zeros $z_n^\mathrm{zeros}$, and the rational function $r(z)$ with respect to design parameters. This proves to be efficient when evaluations of $f(z)$ are costly, but derivatives are available at the expense of a small overhead only, e.g., using algorithmic differentiation. 

\emph{Adding sensitivities to the algorithm.---}In addition to the function values $f_k$, we assume that partial derivatives $\partial f_k/\partial p$ with respect to the design parameters $p \in \mathbb{R}$ at the support points $z_k \in Z$ are available. With the samples $z_k \in Z\backslash\hat{Z}$, we can construct a set of $M-m$ equations,
\begin{equation} \label{eq:system}
\frac{\partial f_k}{\partial p} \approx \sum_{j=1}^{m} \frac{\partial r_k}{\partial f_j} \frac{\partial f_j}{\partial p} + \sum_{j=1}^{m} \frac{\partial r_k}{\partial w_j} \frac{\partial \hat{w}_j}{\partial p},
\end{equation}
for the $m$ unknowns $\partial \hat{w}_j /\partial p$, using $r(z_k) = r_k \approx f_k$. We have $M-m>m$, but only $m-1$ equations are linearly independent. The reason is that Eq.~\eqref{eq:lsq} together with the normalization condition~\eqref{eq:nrm} determines the weights $\hat{w}_j$ up to a global phase $e^{i \varphi}$. 
This factor cancels out in Eq.~\eqref{eq:r}, but adds a degree of freedom to the derivatives of the weights $\partial \hat{w}_j/\partial p$.
From the derivative of Eq.~\eqref{eq:nrm} follows
\begin{equation*}
    \mathrm{Re}\left(\sum_{j=1}^{m} \hat{w}_j^\ast \frac{\partial \hat{w}_j}{\partial p}\right) = 0,
\end{equation*}
but the corresponding imaginary part can be arbitrary and depends on the phase $\varphi$. Indeed, with $\hat{w}_j = w_j e^{i \varphi}$ and using Eq.~\eqref{eq:nrm}, we have
\begin{equation*}
    \mathrm{Im}\left(\sum_{j=1}^{m} \hat{w_j}^\ast \frac{\partial \hat{w_j}}{\partial p}\right) = \mathrm{Im}\left(\sum_{j=1}^{m} w_j^\ast \frac{\partial w_j}{\partial p}\right) + \frac{\partial \varphi}{\partial p}.
\end{equation*}
Since $\varphi$ is an arbitrary real-valued function of the parameter $p$, we can choose $\partial \varphi / \partial p = -\mathrm{Im} \left(\sum_{j=1}^{m} w_j^\ast \partial w_j/\partial p\right)$ and establish the condition
\begin{equation}\label{eq:cond}
    \sum_{j=1}^{m} \hat{w}_j^\ast \frac{\partial \hat{w}_j}{\partial p} = 0. 
\end{equation}
Equations~\eqref{eq:system} and \eqref{eq:cond} fully determine $\partial\hat{w}_j/\partial p$, and we can provide derivatives of poles, zeros, and residues based on $n(z)$ and $d(z)$ as defined in Eq.~\eqref{eq:r},
\begin{equation}
\begin{aligned}
\frac{\partial z_n^\mathrm{pole}}{\partial p} &= \left(\frac{\partial d}{\partial p} \middle/ \frac{\partial d}{\partial z} \right)_{z = z_n^\mathrm{pole}}\\
\frac{\partial a_n}{\partial p} &= \frac{\partial}{\partial p} \left(n \middle/ \frac{\partial d}{\partial z}\right)_{z = z_n^\mathrm{pole}}\\
\frac{\partial z_n^\mathrm{zero}}{\partial p} &= \left(\frac{\partial n}{\partial p} \middle/ \frac{\partial n}{\partial z} \right)_{z = z_n^\mathrm{zero}}.
\end{aligned}
\end{equation}

\emph{Reusing the weights $\hat{w}_j$ to obtain resonance modes.---}If the scalar function values $f_k = \mathcal{L}(\mathbf{f}_k)$ result from a linear map $\mathcal{L} : \mathbb{C}^N \rightarrow \mathbb{C}$, the linearity of $r(z)$ in terms of $\hat{f}_j$ suggests that
\begin{equation*} \label{eq:rv}
    \mathbf{r}(z) = \left. \sum_{j=1}^{m}\frac{\hat{w}_j \hat{\mathbf{f}}_j}{z-\hat z_j}\middle/ d(z) \right.
\end{equation*}
is a good approximation of $\mathbf{f}(z)$. This is of special interest if $\mathbf{f}$ is the solution of the linear system $\mathbf{A}(z) \mathbf{f}(z) = \mathbf{j}(z)$ with a linear operator $\mathbf{A}(z)$ and a source term $\mathbf{j}(z)$. If the scalar function $f(z)$ couples to a given pole $z_n^\mathrm{pole}$, the corresponding residue
\begin{equation} \label{eq:eigenvector}
    \mathbf{a}_n = \sum_{j=1}^m \left[\frac{\hat{w}_j}{z_n^\mathrm{pole}-z_j} \middle/ \frac{\partial d}{\partial z}\left(z_n^\mathrm{pole}\right)\right]\hat{\mathbf{f}}_j
\end{equation}
approximates a solution of the associated nonlinear eigenproblem, i.e., a resonance mode, and, consequently, $\mathbf{A}(z_n^\mathrm{pole}) \, \mathbf{a}_n \approx 0$.

\begin{figure}
\includegraphics{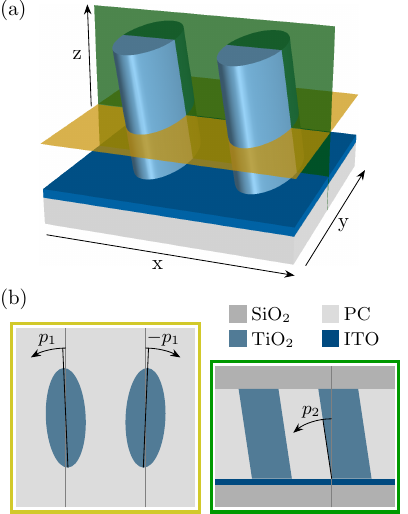}
\caption{\label{fig2}
Chiral resonant metasurface~\cite{Zhang_2022_Science}. (a) Two elliptic titanium dioxide (TiO\textsubscript{2}) bars in a stratified medium constitute the quadratic unit cell with a side length of \qty{360}{\nano\meter}. The positions of the colored planes align with the cross-sections shown below. (b) In-plane and out-of-plane asymmetry parameters $p_1 = \ang{2.5}$ and $p_2 = \ang{8.5}$, respectively. The major and minor axes of the ellipses measure \qty{200}{nm} and \qty{80}{nm} in length, respectively, the latter being also the width of the gap. The bars fit into a \qty{180}{\nano\meter} thick polycarbonate (PC) layer located on a silicon dioxide (SiO\textsubscript{2}) substrate with a \qty{13}{\nano\meter} thick coating of slightly absorbing indium tin oxide (ITO). The top layer consists of polymethyl methacrylate (PMMA), which is modeled with the same constant refractive index of $n=1.45$ as the substrate. A refractive index of $n=1.6$ was adopted for PC, while for TiO\textsubscript{2} and ITO four generalized Drude-Lorentz poles were fitted to empirical data with the AAA algorithm (we refer to the associated data publication~\cite{Betz_SourceCode_AAA} for details). 
}
\end{figure}

\begin{figure*}
\includegraphics{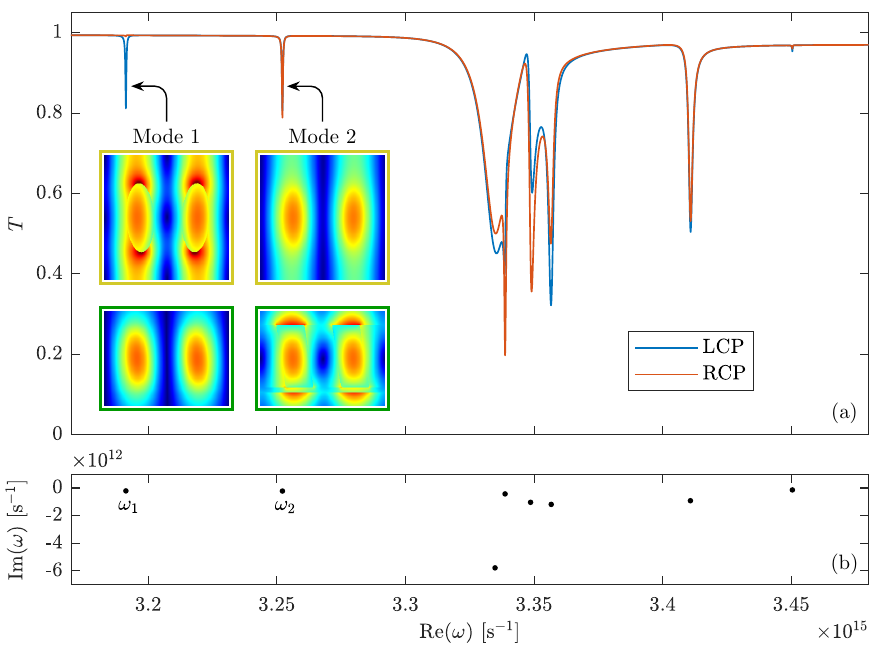}
\caption{\label{fig3}
Transmission and resonances of the metasurface shown in Fig.~\ref{fig2}.
(a) Spectra of normally incident, left and right circularly polarized (LCP and RCP) plane waves. The electric field intensities (a.u.) of the resonance modes at the resonance frequencies $\omega_1$ and $\omega_2$ are displayed in insets where the colors of the frames refer to the cross sections indicated in Fig.~\ref{fig2}. The transmission $T$ is shown for $\mathrm{Im}(\omega) = 0$. (b) Complex frequency plane with computed relevant resonance frequencies $\omega_n$ marked with dots.
}
\end{figure*}

\begin{figure}
\includegraphics{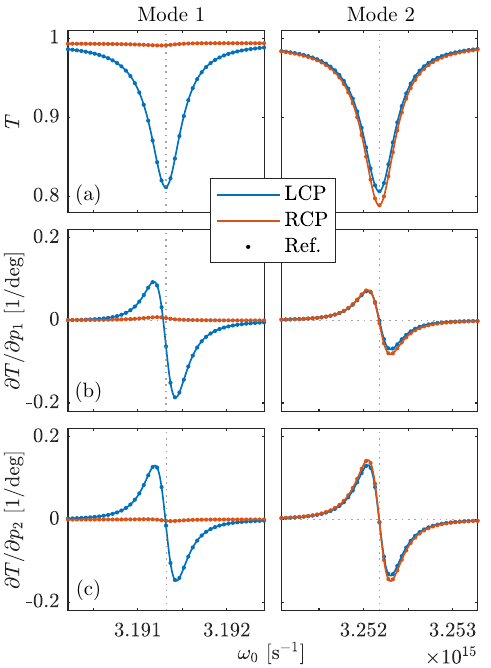}
\caption{\label{fig4}
Transmission spectra and their sensitivities near the resonance modes highlighted in Fig.~\ref{fig3}(a).
(a) Transmission spectra.
(b) Sensitivities with respect to the in-plane asymmetry parameter $p_1$, (c) and with respect to the out-of-plane asymmetry parameter $p_2$.
The real parts of the resonance frequencies $\omega_1$ and $\omega_2$ are indicated by vertical dotted lines.
Reference solutions are shown for validation purpose and are marked with dots. Each reference point corresponds to the solution of a scattering problem at the associated real-valued frequency~$\omega_0$. 
}
\end{figure}

\emph{Application.---}The effectiveness of the AAA algorithm is demonstrated with a chiral metasurface~\cite{Zhang_2022_Science}. The structure of its unit cell is shown in Fig.~\ref{fig2}(a) and (b). The focus is on the transmission $T(\omega_0)$ resulting from a circularly polarized plane wave propagating in the positive z direction with the real-valued angular frequency $\omega_0 \in \mathbb{R}$. The corresponding time-harmonic scattering problems are mathematically described by the second-order Maxwell's equation
\begin{equation}
	\nabla \times \mu^{-1} 
	\nabla \times \mathbf{E} -
	\omega_0^2\epsilon \mathbf{E}  = 
	i\omega_0\mathbf{J}. \label{eq:Maxwell}
\end{equation}
From its solution, the electric field strength $\mathbf{E}(\mathbf{r},\omega_0) \in \mathbb{C}^3$, all electromagnetic quantities of interest can be derived. In Eq.~\eqref{eq:Maxwell}, the source is modeled as an electric current density $\mathbf{J}(\mathbf{r})\in \mathbb{C}^3$, the material properties are integrated into the model through the permittivity $\epsilon(\mathbf{r},\omega_0)$ and permeability $\mu(\mathbf{r},\omega_0)$ tensors, and $\mathbf{r} \in \mathbb{R}^3$ is the position. 

In principle, the AAA algorithm can build rational approximations exclusively from samples along the real axis. However, the number of sample points can be reduced if data is available at complex evaluation points. 
Eq.~\eqref{eq:Maxwell} can be solved for $\mathbf{E}(\mathbf{r},\omega)$ at complex frequencies $\omega \in \mathbb{C}$ and due to the periodicity of the structure, we compute the transmission $T(\omega)$ based on its Fourier coefficients $\mathbf{E}_\mathbf{k}(\omega)$ associated with a discrete set of wave vectors $\mathbf{k}$. Special care is required since $T(\omega)$ is quadratic in the Fourier coefficients. For more information on the analytic continuation of quadratic quantities, see Ref.~\cite{Binkowski_2020_PRB}.

We address Eq.~\eqref{eq:Maxwell} for different frequencies using the finite element method (FEM) implemented in the software package JCMsuite. The accuracy of the discrete model is verified by assessing reflection, transmission, and absorption at 241 evenly spaced points within the specified range. Energy conservation requires that the discrepancy between their sum and the energy entering the system is zero. 
The numerical discretization is chosen such that
the maximum relative error is less than $3 \times 10^{-5}$.

Solving Eq.~\eqref{eq:Maxwell} numerically is computationally expensive, and the number of support points for the AAA algorithm should be as small as possible. This was one of the motivations for the recently published adaptive discretization scheme for the AAA algorithm~\cite{Driscoll_2024}. The so-called continuous AAA needs to be versatile enough to generalize to many different problem types, e.g., functions with essential singularities. For applications in physics, assumptions can usually be made about the systems. Here, we anticipate the predominance of simple poles in the vicinity of the real frequency axis. We expect the corresponding resonance frequencies to converge quickly and the emergence of some less stable poles that provide an interpolation of the background.

Our sampling approach involves two phases. Initially, a preliminary estimate of the relevant resonance frequencies $\omega_n$ is inferred from $M+1$ evenly spaced real-valued frequencies. We consider a resonance frequency relevant if its real component falls within the specified range and its quality factor is higher than a selected value, i.e., the imaginary part is less than \qty{10}{\percent} of the largest real-valued frequency in magnitude. The approximate error of a given resonance frequency $\omega_n$ is evaluated by comparing the results based on all the available support points with the results after eliminating the $\hat{z}_j$ closest to $\omega_n$. If this error is larger than a threshold, we repeat the first phase with $2M+1$ sample points. We choose the distance between neighboring sample points as the threshold value. In the final phase, we add $2N$ data points in the vicinity of the $N$ relevant resonance frequencies. 

Hereafter, rational approximations are constructed based on $M+1=31$ real-valued sampling points, plus 16 and 32 additional points for quantities linear and quadratic in the electric field, respectively, pertaining to the $N=8$ relevant resonance frequencies. 
Comparison with $241$ reference points yields a maximum absolute error of the reconstructed spectra, including sensitivities, below $10^{-6}$, in contrast to an error greater than $10^{-1}$ after the first phase. The relative errors of the resonance frequencies compared to reference solutions of the nonlinear eigenproblem obtained with the Arnoldi method are of the order $10^{-5}$ after the first phase and drop below $10^{-7}$ in the second phase.

Figure~\ref{fig3}(a) shows the transmission of the metasurface, together with the electric field intensity plots of two selected resonance modes $\mathbf{E}_{n}(\mathbf{r})$, which are solutions of Eq.~\eqref{eq:Maxwell} when the source is set to zero, i.e., of the nonlinear eigenproblem. The underlying resonance frequencies in the complex frequency plane are depicted in Fig.~\ref{fig3}(b). Mode 1 shows the property for which the authors aim in Ref.~\cite{Zhang_2022_Science}. The authors associate the field distribution pattern with the fields generated by two dipoles and establish a criterion to determine the asymmetry parameters $p_1$ (in plane) and $p_2$ (out of plane) to obtain an optimal coupling of Mode 1 to the LCP signal while minimizing its effect on the RCP signal.
For the development of such a simplified model, it is crucial to know the field distributions of the resonance modes. Since the Fourier transform is a linear operation, the poles of the coefficient $\mathbf{E}_\mathbf{k}(\omega)$ correspond to the resonances of the electric field $\mathbf{E}(\mathbf{r},\omega)$. Inserting the weights $\hat{\omega}_j$ of its rational approximation in Eq.~\eqref{eq:eigenvector}, we can determine the resonance modes $\mathbf{E}_n(\mathbf{r})$. 
This method works remarkably well.
In fact, compared to reference solutions obtained with the Arnoldi method, the error is $\mathrm{err}(\mathbf{E}_n) < 10^{-8}$, where $\mathrm{err}(\mathbf{E}_n) = \left\Vert \mathbf{E}_n-\mathbf{E}_n^{\text{ref}}\right \Vert/\left\Vert\mathbf{E}_n\right \Vert$. The norm $\Vert \cdot \Vert$ is defined as the volume integral of the electric field energy density over the computational domain.

If a simplified model is not available to provide optimal parameter combinations, optimization methods based on the full problem must be utilized and can be accelerated if derivative information is available. Furthermore, the insights they provide can, e.g., simplify the identification of suitable design parameters. Figure~\ref{fig4} displays the spectra in a close vicinity of the resonance frequencies $\omega_1$ and $\omega_2$, which were introduced in Fig.~\ref{fig3}(b) and whose real parts are indicated by horizontal dotted lines. The transmission characteristics of Mode 1 displayed in Fig.~\ref{fig4}(a) are close to optimal. Yet, with the derivative information in Fig.~\ref{fig4}(b) and (c) at hand, we can infer that increasing $p_1$ would further decrease the transmission of LCP and increase the transmission of RCP, while a change of $p_2$ would primarily shift the dip in the LCP signal. In contrast to Mode 1, the responses of Mode 2 to the different polarizations are very similar. To demonstrate the accuracy of the results, we have added the reference solutions, which correspond to scattering solutions on the real-frequency axis. It can be observed that, even in the close vicinity of the resonance frequencies, the rational approximations are very accurate. Such accurate and efficient approximations of sensitivities are particularly useful when the systems to be optimized are characterized by many design parameters.

\emph{Conclusion.---}We presented a theoretical framework for the investigation of resonant photonic systems. The AAA algorithm is used to accurately approximate the optical response of the systems. The resulting rational approximations are based on the response at adaptively chosen sample points in the complex frequency plane. We extended the AAA algorithm to compute also sensitivties in an efficient way and we applied the algorithm to extract the underlying resonance modes. 
The high accuracy of the rational approximations is demonstrated by a comparison to fine sampled reference solutions at real-valued frequencies.
All methods presented can be transferred to other resonance-based problem classes, e.g., in quantum mechanics or acoustics.
We applied the framework to a chiral metasurface. Resonances with and without strong chiroptical response were detected and investigated.

Despite its impact on other fields of science and engineering, to the best of our knowledge, the AAA algorithm has not yet been applied to problem classes in the field of photonics. We expect that the presented approach will allow to reliably investigate and optimize setups which were relatively difficult to tackle with numerical methods so far, e.g., due to high-quality factor resonances, strongly multimodal behavior, complex material dispersion, or branch cuts in the complex frequency plane. In particular, we expect that the framework will be useful to investigate different kinds of singularities in photonics, such as poles and reflection zeros of metasurfaces~\cite{Elsawy_2023}, or exceptional points of nanoscale systems~\cite{Li_2023}.

\emph{Data availability.}---Supplementary data tables and source code for the numerical experiments
for this work can be found in the open access data publication~\cite{Betz_SourceCode_AAA}.

\emph{Acknowledgments.}---We acknowledge funding
by the Deutsche Forschungsgemeinschaft (DFG, German Research Foundation) 
under Germany's Excellence Strategy - The Berlin Mathematics Research
Center MATH+ (EXC-2046/1, project ID: 390685689) and
by the German Federal Ministry of Education and Research
(BMBF Forschungscampus MODAL, project 05M20ZBM).

\end{document}